\documentclass{mem}
\usepackage{natbib}\usepackage{txfonts}\usepackage{balance}
\usepackage{graphicx}
\usepackage[a4paper,breaklinks,dvipdfm]{hyperref}
\idline{00}{000}
\begin{document}
\def\teff{$T\rm_{eff }$}
\def\kms{$\mathrm {km s}^{-1}$}

\title{
Probing brown dwarf formation mechanisms with {\it Gaia}
}

   \subtitle{}

\author{
Richard J. Parker\inst{} 
          }

  \offprints{R. Parker}

\institute{
Astrophysics Research Institute,
Liverpool John Moores University,
146 Brownlow Hill,
Liverpool,
L3 5RF,
UK.
\email{R.J.Parker@ljmu.ac.uk}
}

\authorrunning{Parker }

\titlerunning{BD formation with {\it Gaia}}

\abstract{
One of the fundamental questions in star formation is whether or not
brown dwarfs form in the same way as stars, or more like giant
planets. If their formation scenarios are different, we might expect
brown dwarfs to have a different spatial distribution to stars in
nearby star-forming regions. In this contribution, we discuss methods to look
for differences in their spatial distributions and show that in the
only nearby star-forming region with a significantly different spatial
distribution (the Orion Nebula Cluster), this is likely due to dynamical evolution. We
then present a method for unravelling the past dynamical
history of a star-forming region, and show that in tandem with {\it
  Gaia}, we will be able to discern whether observed differences are
due to distinct formation mechanisms for brown dwarfs compared to stars.  
}
\maketitle{}

\section{Introduction}

One of the outstanding questions in star formation is whether brown
dwarfs (BDs) form more like stars, or more like giant planets
\citep{Chabrier14}. One way of addressing this question is to compare
the spatial and velocity distributions of stars and (BDs) in star-forming regions; any differences could indicate that the formation 
mechanisms are different.

Whilst this may appear to be a straightforward question to address, two issues 
complicate the picture. Firstly, methods for comparing the spatial 
distributions of objects are diverse and range from comparing the
slopes of mass spectra as a function of distance from the centre of a
star-forming region, to measuring the relative concentration or local
surface density around BDs compared to stars. The latter two methods of
measuring mass segregation can give contradictory results, especially if
the morphology of a star-forming region is complex.

 Secondly, early
dynamical evolution in star-forming regions is known to alter the
spatial distributions of stars, and could be responsible for any
observed difference is the spatial distributions of stars and BDs. In
this scenario, we would need to know the amount of dynamical evolution
that has occurred in the past to assess whether any differences in
spatial distributions are a relic of the star and BD formation
process(es) in the star-forming region in question.

\section{Spatial distributions of stars and BDs}

\subsection{Quantifying spatial differences}

In this section, we discuss the spatial distributions of stars and BDs in three star-forming regions with different densities and using two different
measures of mass segregation. Classically, mass segregation is the
over-concentration of massive stars in a cluster, and can be
indicative of ongoing energy equipartition in a stellar system. 

We determine the relative spatial distribution of stars to BDs using the
minimum spanning tree (MST) $\Lambda_{\rm MSR}$ ratio
\citep{Allison09a}, which compares the MST length of random stars in a
region to the MST length of a chosen subset of stars (or in this case,
BDs). We then compare the median local surface density around the BDs 
to the median local surface density of all objects in the region, the
$\Sigma_{\rm LDR}$ ratio \citep{Maschberger11,Parker14b}.

\subsection{Spatial distributions in Taurus, $\rho$~Oph and the ONC}

We measure $\Lambda_{\rm MSR}$ and $\Sigma_{\rm LDR}$ in the
low-density ($\tilde{\Sigma} = $ 5\,stars\,pc$^{-2}$) Taurus association,
medium density ($\tilde{\Sigma} = $ 75\,stars\,pc$^{-2}$) $\rho$~Oph
region, and the higher density Orion Nebula Cluster (ONC, $\tilde{\Sigma} = $ 1000\,stars\,pc$^{-2}$).

In Taurus there are hints that the BDs have a more sparse spatial
distribution compared to stars, but the differences are not
particularly significant \citep{Parker11b}. In $\rho$~Oph, there is no
preferred spatial distribution \citep{Parker12c}. In the ONC, however,
the BDs appear to be more sparse than the higher mass stars. This
could be due to dynamical evolution; the most massive stars in the ONC
are mass segregated \citep{Allison09a}, although the timescale to mass
segregate down to beyond the hydrogen-burning limit (and hence reach
full energy equipartition) is probably of order a Hubble time (and
certainly orders of magnitude longer than the age of the ONC). 
However, the destruction of primordial binaries which contain BDs may
result in the ejection of BDs at greater velocities (and hence
traveling further and becoming more distributed) than the low-mass stars.

\section{Dynamical evolution}

Can the observed spatial differences between stars and BDs in some
star forming regions be explained by dynamical interactions? In the
ONC, the mass segregation of the most massive stars can be explained in
the cluster formed from the collapse of a subvirial, substructured
star forming region \citep{Allison09b}. Many star forming regions are
observed to be subvirial \citep[e.g.][]{Peretto06}, and substructure
is ubiquitous \citep[e.g.][]{Cartwright04,Arzoumanian11,Gouliermis14}.

However, it is unclear whether these initial conditions also lead to
the preferential ejection of BDs over low mass stars. It is important
to note that comparing the spatial distribution of BDs to massive
stars that are already mass segregated will in most cases give the
erroneous result that the the BDs are `inversely mass segregated'.

In \citet{Parker14c} we followed the evolution of the spatial
distribution of brown dwarfs compared to low mass stars in
$N$-body simulations of substructured, subvirial regions undergoing cool-collapse
\citep{Allison09b}. We measured the $\Lambda_{\rm MSR}$ ratio, which
compares the overall spatial concentration for BDs compared to stars;
and the $\Sigma_{\rm LDR}$ ratio, which compares the local density around
BDs compared to stars. We compared this to the ratio of stars to brown
dwarfs as a function of distance from the cluster centre,
$\mathcal{R}_{\rm RSS}$ \citep[e.g.][]{Andersen11}; an observed decrease in this ratio from
the centre would indicate that more BDs are on the outskirts. 

The $N$-body simulations contained 1500 stars and BDs drawn from a
\citet{Maschberger13} IMF, and binary properties as observed in the
Galactic field. The initial spatial distributions of the stars and BDs
were identical to each other. We then evolved the star forming regions for 10\,Myr,
during which time a smooth, centrally concentrated cluster similar to
the ONC formed.

In \citet{Parker14c} we found that in a suite of twenty identical simulations (differing
only in the random number seed used to generate the initial
conditions), 6/20 show the BDs to be more spread out according to all
three diagnostics: $\Lambda_{\rm MSR}$, $\Sigma_{\rm LDR}$ and
$\mathcal{R}_{\rm RSS}$. In a larger fraction of simulations, the BDs
appeared more spread out in two of the three diagnostics. However, in many simulations these measures
were transient, and only presented themselves as being significant for
a small fraction of the full dynamical evolution. 

Based on these simulations, we conclude in \citet{Parker14c} that the
observed difference in spatial distribution of BDs compared to stars
could be due to the dynamical evolution, rather than differences in
the outcome of the star/BD formation process(es). (We also note that
more complete observations would be highly desirable in order to
ascertain whether the observed differences are actually real.)     

\section{Dynamical histories of star forming regions with {\it Gaia}}

Whilst we have shown that dynamical interactions could be responsible
for the observed differences in spatial distributions between stars
and brown dwarfs, we often have very little information on the
previous evolution of the star forming region or cluster, which makes
in difficult to conclusively attribute spatial differences to
dynamics.

However, if we also utilise other measures of spatial structure, and
ultimately kinematical information from {\it Gaia} and its associated
ground-based spectroscopic surveys, we will be able to calibrate
suites of simulations to observed diagnostics, and determine the
initial kinematic and spatial distributions of these regions. This
would allow us to assess whether a region has been so dense in the
past that dynamics could have been responsible for any observed
difference between stars and BDs. 

To do this, we combine measures of mass segregation $\Lambda_{\rm MSR}$ and local surface
density $\Sigma_{\rm LDR}$ for the {\it most massive stars in the
  region, rather than the BDs}  with the spatial structure of the
region, the $\mathcal{Q}$-parameter
\citep{Cartwright04,Cartwright09a}. Regions with substructure have 
$\mathcal{Q} < 0.8$, and centrally concentrated regions without
substructure have  $\mathcal{Q} > 0.8$. Dynamical interactions rapidly
erase substructure \citep{Parker12d,Parker14b} so any region with
$\mathcal{Q} >1$ has likely undergone significant dynamical evolution
\citep{Parker12d}.

Furthermore, if a region has a high local density initially
($>100$\,stars\,pc$^{-2}$) then the $\Sigma_{\rm LDR}$ ratio becomes
significantly higher than unity for the massive stars, irrespective of
the initial velocity dispersion. On the other
hand, $\Lambda_{\rm MSR}$ only becomes significantly greater than
unity in subvirial collapsing regions.  

Combining all three measures, a subvirial region which collapses to
form a dense cluster will have  $\mathcal{Q} >1$ and $\Sigma_{\rm LDR}
> 1$ and $\Lambda_{\rm MSR} >1$, whereas a supervirial region which has expanded will retain some
stucture $\mathcal{Q} < 0.8$ and have $\Sigma_{\rm LDR} > 1$ and
$\Lambda_{\rm MSR} \sim 1$. It is the former scenario which has
likely produced the ONC.

A more quiescent (i.e.\,\,low local densities) region, even if in the process of collapsing, will
have a $\mathcal{Q}$-parameter $< 1$, and $\Sigma_{\rm LDR} \sim 1$
and $\Lambda_ {\rm MSR} \sim 1$. In such a region, if the distribution
of BDs is different to stars, then it is unlikely that dynamical
interactions alone are responsible and one can conclude that star
formation and brown dwarf formation are different.   

The {\it Gaia} satellite, and its associated ground-based
spectroscopic surveys will be able to deliver kinematical information
on stars on the outskirts of unobscured star clusters, which can then
be compared to tailor-made numerical simulations. 

In a preliminary study, \citet{Allison12} compared the dynamical
evolution of star-forming regions that formed clusters, but with subtle
differences in the initial conditions. In one set of simulations, the
star-forming regions were subvirial, with a high amount of
substructure, and in the other set the regions were in virial
equilibrium and the stars were arranged in an almost-uniform spherical
distribution. 

\citet{Allison12} showed that the more substructured, subvirial
regions ejected more stars into an outer halo, and those stars were
ejected with higher velocities, than the virialised, smoother
regions. At first glance, it appears that {\it Gaia} would be able to
readily distinguish between these initial conditions. However, the
addition of primordial binaries causes this result to be diluted,
because the binaries effectively act as extra substructure in the
simulations. 

A more in-depth analysis of the velocity space in these simulations is
required to fully understand the capabilities of {\it Gaia} in respect
of discerning the initial conditions of star formation in exposed
(open) 
clusters. However, any information that is present in the velocity
space will compliment the spatial distribution analyses discussed in
this article.

\section{Conclusions}

Current observations of star-forming regions have largely shown that
the spatial distributions of stars and brown dwarfs (BDs) are very similar,
if not indistinguishable. In several regions, most notably the Orion
Nebula Cluster, the BDs have a slightly more sparse spatial
distribution. However, this can be explained by the star-forming
region have undergone dynamical evolution, which scatters the BDs to
the outskirts. 

Using the full 2D spatial information for {\it all} members in a
region can place limits
on the amount of dynamical evolution which has taken place, allowing
us to assess whether any observed difference between the stars and BDs
is likely due to their formation mechanisms being distinct. Further
kinematical information from {\it Gaia} and its associated
ground-based spectroscopic surveys will also place strong constraints
on the dynamical histories of star-forming regions, and by extension,
the brown dwarfs within.
 
\begin{acknowledgements}
I thank Ricky Smart and the LOC/SOC for giving me the opportunity to
present this work at the conference. I also thank Yifat Dzigan and
Esther Buenzli for extracting us from the conference dinner kareoke in the
nick of time. 
\end{acknowledgements}

\bibliographystyle{aa}
\bibliography{Parker_Torino}

\end{document}